# Optimal exit decision of venture capital under time-inconsistent preferences

Yanzhao Li[a], Ju'e Guo[a,b], Yongwu Li[c*], Xu Zhang[a,b]

[a] School of Management, Xi'an Jiaotong University, Xi'an, China
[b] Research Center of Chinese Management, Xi'an Jiaotong University, Xi'an, China
[c] School of Economics and Management, Beijing University of Technology, Beijing, China
E-mail: liyw@bjut.edu.cn (Yongwu Li, corresponding author)

February 22, 2020

Abstract

This paper proposes two kinds of time-inconsistent preferences (i.e. time flow inconsistency and critical time point inconsistency) to further advance the research on the exit decision of venture capital. Time-inconsistent preference, different from time-consistent preference, assumes that decision makers prefer recent returns rather than future returns. Based on venture capitalists' understanding of future preferences, we consider four types of venture capitalists, namely time-consistent venture capitalists, venture capitalists who only realize critical time point inconsistency, naïve venture capitalists and sophisticated venture capitalists, of which the latter three are time-inconsistent. All types of time-inconsistent venture capitalists are aware of critical time point inconsistency. Naive venture capitalists misunderstand time flow inconsistency while sophisticated ones understand it correctly. We propose an optimal exit timing of venture capital model. Then we derive and compare the above four types of venture capitalists' exit thresholds. The main results are as follows: (1) all types of time-inconsistent venture capitalists tend to exit earlier than time-consistent venture capitalists. (2) The longer the expire date are, the more likely venture capitalists are to delay the exit, but the delay degree decreases successively (venture capitalists who only realize critical time point inconsistency > naïve venture capitalists > sophisticated venture capitalists).

Acknowledgments

The first and second authors acknowledge funding from the Major Program of the National Social Science Foundation of China (No.17ZDA083). The corresponding author acknowledges funding from Beijing Natural Science Foundation (No. 9192001) and National Natural Science Foundation of China (No. 71932002). The fourth author acknowledges funding from the Soft Science Research Program of Shaanxi Province, China (No.2016KRM064).

1. Introduction

Venture capital (VC) provides the imperative capital for the development of entrepreneurial enterprises. In addition, venture capital, due to its unique mode of operation, plays the role of coping with risks, facilitating venture success and nurturing high-tech industries around the world, especially in transition economies like China. According to KPMG's quarterly report on venture capital trends named "Venture Pulse", both Asian and global venture capital transactions increased by more than 40% in 2018, reaching a record of $93.5 billion and $254.7 billion US dollars, respectively. Among them, Chinese venture capital transaction volume reached a record of 70.5 billion US dollars, an increase of 52.9% compared with 46.1 billion US dollars in 2017.

The exit, the divestment of the company from the VC's portfolio, plays a crucial role since it achieves the sale of the shares and obtains value-added returns, and thus determines the final potential gain of the VC fund. In this case, exit earnings are a signal of the VC fund's quality (Cumming 2010). In particular, for a start-up, not prestigious VC fund, it is strictly the only quality signal. And the quality signal decides the success of follow-up fund-raising (Cumming 2010). In general, researchers have devoted a great deal of attention to the optimal exit decision of VC, including exit route and exit timing, such as Hellmann (2006), Giot and Schwienbacher (2007), Cumming and Johan (2008), Bienz and Walz (2010), Smith *et al.* (2011), Li and Chi (2013), Arcot (2014), Bock and Schmidt (2015).

The two main exit routes of VC include initial public offering (IPO) and trade sale (merger & acquisition, for short, we call it M&A later). For almost two decades, trade sale has always been the dominant exit route in America and Europe due to transaction flexibility (Bienz & Walz 2010). For example, IPOs have represented only 10-20% of all exits in the US VC industry, while 80%-90% of firms have been sold via M&A since 2001. Meanwhile, this similar situation has also occurred in China despite the launch of SSE STAR MARKET (known as Nasdaq of China) promotes lots of companys to IPOs in 2019. For this reason, venture capitalists are gradually focusing on trade sale. Nishihara (2017) points out that the previous literature focused on IPO issues while igoring alternative choices. Additionally, stock market conditions are more crucial for companies supported by VC than decision makers' preference as their flexibility in IPO timing is limited. Hence our paper focuses on trade sale, and derives the optimal exit timing of the VC using real options.

It is worth noting that time preferences and beliefs of venture capitalists affect the optimal exit timing of VC. Both the agent's time preferences for returns and VC's organizational structure cause venture capitalists to have time-inconsistent preferences.

Time-inconsistent preference is different from time-consistent preference, and decision makers prefer recent returns rather than future returns. Lots of experimental studies in behavioral finance suggest that the assumption of time-inconsistent preference is realistic, such as Strotz (1955), Thaler (1981), Loewenstein and Prelec

(1992). Time inconsistency is also reflected in the fact that the discount rate for returns is decreasing with time. Laibson (1997) proposes a hyperbolic discount function instead of an exponential function to correctly reflect the change in the discount rate. Grenadier and Wang (2007) extend the real options to model the investment timing decisions of entrepreneurs with time-inconsistent preferences firstly, and they investigate cases of monopoly entrepreneurs and competitive equilibrium. Tian (2016) explores the leverage choices of entrepreneurs with time-inconsistent preferences in financing, and that study shows that naïve entrepreneurs are more likely to choose higher leverages, while sophisticated ones always choose lower leverages. Chen *et al.* (2016) consider an optimal dividend-financing problem for a company whose managers have time-inconsistent preferences, and their results show that the manager with time-inconsistent preferences tends to pay out dividends earlier. Guo *et al.* (2018) analyze the impact of decision makers' time-inconsistent preferences on investment timing for rail transit and provide similar conclusions with Grenadier and Wang (2007). However, little of this research has attempted to investigate the effect of time-inconsistent preference of venture capitalists.

However, time-inconsistent preference literature mentioned above is caused by the agent's time preference, which is only related to the time flow. But beyond that, VC's organizational structure, specifically the finite lifespan of VC funds is also the source of time-inconsistent preference for venture capitalists.

Limited partnership has been the main organizational form of VC funds for the past three decades. It is designed with the finite lifespan: approximate 10 years in America(e.g., Gompers and Lerner (1999), Kandel *et al.* (2011)), 5 even 3 years in China, occasionally with a possible extension of 2-3 years. Several attempts have been made to explore the impact of the finite lifespan of VC funds on the decision-making of venture capitalists. Kandel *et al.* (2011) proves that termination of all unfinished projects at the fund's maturity leads to suboptimal decisions during later stages of investment by theoretical arguments, and proposes several possible contractual remedies. Arcot *et al.* (2015) investigate whether secondary buyouts are value-maximizing, or reflect opportunistic behavior. They show that funds under pressure (ie. the finite lifespan) engage more in secondary buyout. Actually, the discount rate of venture capitalists drop rapidly after VC funds expiring, which is the embodiment of time-inconsistent preference.

In order to distinguish the two kinds of time inconsistency mentioned above, our paper divides time inconsistency into 'time flow inconsistency' and 'critical time point inconsistency'. During the investment period of VC funds (as shown in Fig.1), venture capitalists have time flow inconsistency, which only depends on venture capitalists' trade-offs between time and returns. The finite lifespan of VC funds forces venture capitalists to sell all the projects before the expire date (Kandel *et al.* 2011). This pressure makes venture capitalists to have a much lower utility for the earnings of VC funds after the expire date. Although there is the extension period (see Fig.1) for most VC funds, funds' investors can observe the expected return of venture capitalists. In consequence, venture capitalists have a strong incentive to exit before the expire date. In summary, venture capitalists have critical time point inconsistency before and after the expire date. Obviously, the degrees of these two kinds of time inconsistency is different.

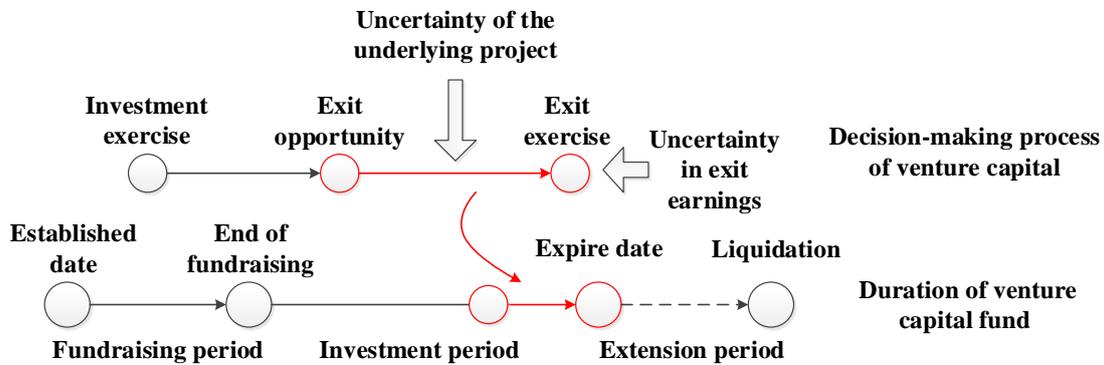

Fig.1 The schematic of this paper

Referring to the previous literature, our paper examines the optimal exit timing of (1) time-consistent venture capitalists; (2) venture capitalists who only realize critical time point inconsistency; (3) naïve venture capitalists and (4) sophisticated venture capitalists. Following but different from the previous literature, all types of time-inconsistent venture capitalists are aware of critical time point inconsistency, but naive venture capitalists misunderstand time flow inconsistency, who assume that future selves (an explanation in Section 2) act according to preferences of the current self during the investment period of VC funds, while sophisticated venture capitalists assume that future selves choose strategies that are optimal for themselves.

Giving exogenously the profit flow of the invested company, our paper proposes an optimal venture capital exit option model under trade sale. Furthermore, the optimal exit timings of the above four types of venture capitalists are derived, and the sensitivity analysis of the key factors is carried out. The results are summarized as follows: (1) all types of time-inconsistent venture capitalists tend to exit earlier than the time-consistent venture capitalists; (2) the longer the expire date are, the more likely time-inconsistent venture capitalists are to delay the exit, but the delay degrees decrease successively (venture capitalists who only realize critical time point inconsistency > naïve venture capitalists > sophisticated venture capitalists). (3) the exit threshold of venture capitalists is positively correlated with exit cost and volatility of the invested company's profit, while it is negatively correlated with negotiation ability of venture capitalists, fixed income in M&A, equity ratio of the VC, and expected growth rate of the invested company's profit.

The objective of this paper is to examine venture capital funds' (particularly start-up, not prestigious VC funds') exit timing by trade sale under time-inconsistent preferences. To our best knowledge, no attempt has been made to consider the exit decision of venture capital under time-inconsistent preferences. We unify two kinds of time inconsistency in mathematics, respectively from time flow inconsistency mentioned in the previous literature, and critical time point inconsistency caused by the finite lifespan of VC funds. The most relevant literature in research method is Grenadier and Wang (2007), who consider entrepreneures' investment timing decisions under time flow inconsistency. A large number of literatures use real options approach to study investment problems in various fields (for example, wind power generation (Muñoz *et al.* 2011), venture capital (Ko *et al.* 2011), mobile telecommunications networks (Franklin 2015)), mainly focusing on the timing of entry. In order to simplify the model, such literature generally ignores "time to build" (Marmer & Slade 2018). However, the exit of investment is naturally less affected by "time to build" than the entry, so it's more suitable to be described by real options approach.

The remainder of the paper is organized as follows. Section 2 provides the model setup including venture capitalists' time preferences and the exit opportunity under trade sale. Section 3 derives the optimal exit timing of time-consistent venture capitalists, venture capitalists who only realize critical time point inconsistency, naïve venture capitalists and sophisticated venture capitalists, respectively. Section 4 gives the sensitivity analyses of model parameters and the corresponding analysis. Section 5 concludes this paper.

2. Model setup

2.1 Venture capitalists' time preferences

We follow Chen *et al.* (2017), Harris and Laibson (2012) and Grenadier and Wang (2007) to describe time preferences of venture capitalists by using a continuous-time quasi-hyperbolic discount function. The venture capitalist is described as a finite number of selves with a random lifespan, and each self represents the current stage of the venture capitalist to exercise the exit decision, while considering the utility of the future selves exercise decision. As shown in Fig.2, we call $t_0$ the start time moment, which signals the birth of self 0 and means an exit option emerges for the venture capitalist to sell the share of the invested company. Let $T_E$ denote the duration from $t_0$ to the expiry date of the VC fund. We assume that $T_E$ is exponentially distributed with parameter $\lambda_E$. In fact, time uncertainty of $T_E$ comes from the uncertainty of the start time moment $t_0$. However, if the moment $t_0$ is assumed to be the reference point in order to facilitate comparison, the arrival of the expiry date of the VC fund is modeled as a Poisson process with intensity $\lambda_E$. Let $t_n$ be the birth time of self $n$ and the death time of self $n-1$ $(n=1,2,3,...,E-1,E)$. The lifespan $T_n = t_{n+1} - t_n$ for self $n$ is assumed to be exponentially distributed with parameter $\lambda_f$. If instead the expiry date of the VC fund arrives before the next self which is generated by time flow inconsistency, the next self is changed to self $E$ that generated by critical time point inconsistency. The duration of self $E-1$ is assumed to be exponentially distributed with parameter $\lambda_p$ (including $\lambda_{pN}$ and $\lambda_{pS}$). We noted that $1/\lambda$ represents the interarrival time of selves because $\lambda$ is the arrival intensity of the

Poisson process, then $E\left[(E-1)/\lambda_f + 1/\lambda_p\right] = E\left[1/\lambda_E\right]$.

We use $D_n(t,s)$ to denote the inter-temporal discount function of self $n$, which gives self $n$'s discounted value at time $t$ of one dollar received at the future time $s$. For payoffs obtained in the duration of self $n$, self $n$ uses a standard discount function $e^{-\rho(s-t)}$, where the discount rate $\rho > 0$. For payoffs obtained after the death of the current self, self $n$ uses the discount function $\delta e^{-\rho(s-t)}$, multiplying a reduction factor $\delta$ based on the standard discount function. After the arrival of self $n+1$, the venture capitalist uses the updated discount function $D_{n+1}(t,s)$ for evaluation. It's worth mentioning that the different reduction factor $\delta_f$ for self $n(n=0 \sim E-2)$ and $\delta_p$ for self $E-1$, obviously, $\delta_f > \delta_p$.

Then

$$D_n = \begin{cases} e^{-\rho(s-t)} & \text{if } s \in [t_n, t_{n+1}) \\ \delta e^{-\rho(s-t)} & \text{if } s \in [t_{n+1}, \infty) \end{cases} \qquad (1)$$

for $s > t$ and $t \in [t_n, t_{n+1})$.

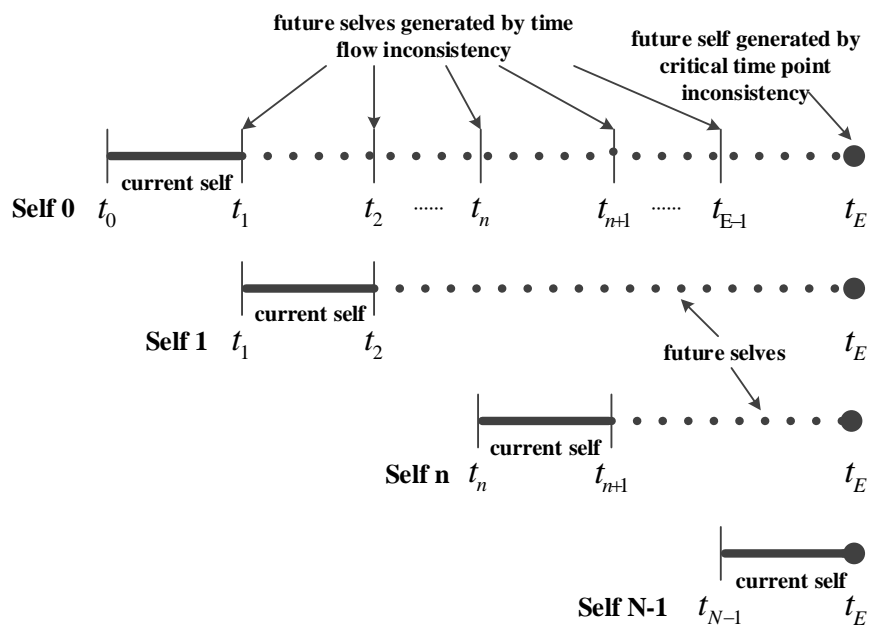

Fig.2 Two kinds of time inconsistency

2.2 The exit opportunity

Consider that the venture capitalist has an exit opportunity for the invested company by trade sale. The invested company has uncertain development prospects. Let $P_t^T$ denotes the profit flow of the company at time $t$. We suppose that the company profit is given by the geometric Brownian motion process:

$$dP_t^T = \alpha P_t^T dt + \sigma P_t^T dB_t, \ t \geq 0 \qquad (2)$$

Where $dB_t$ is the increment of a standard Wiener process, $\alpha$ is the expected growth rate of profit, and $\sigma$ is the volatility of profit.

Following Thijssen (2008), we assume that the profit $P_t^T$ of the invested company (the acquired company) consists of a deterministic part, denoted by $Q^T$, and a stochastic component, denoted by $X_t$. The stochastic shock is assumed to be multiplicative, that is $P_t^T = Q^T X_t$. In the same way, for the acquiring company and the merged company, there are $P_t^A = Q^A X_t$ and $P_t^M = Q^M X_t$. The deterministic component can be thought as resulting from competition in the product market. The stochastic component indicates uncertainty. We suppose that the stochastic component evolves as the geometric Brownian motion process:

$$dX_t = \mu X_t dt + \sigma X_t dB_t, \ t \geq 0, \ X_0 = x \tag{3}$$

In order to maintain the convenience of derivation without loss of generality, it is assumed that the acquiring firm and the venture capitalist hold the same discount rate in this M&A. The payment of dividends, provided by the invested company, is far less than the payoff from selling the shares they hold in the invested company for the VC. Therefore we assume that trade sale is the only way to make a return on the investment. So even though the profit of the invested company is flow over time, the time preference of the venture capitalist does not affect the calculation of the expected present value of the profit flow generated by the invested company.

Gao et al. (2013) empirically point out that the acquiring company can expand business more efficiently through achieving economies of scale and scope. In this paper, this positive effect is specifically characterized as a synergy that the resulting deterministic profit is larger than the sum of the deterministic profits of the constituent firms, which is $Q^M > Q^A + Q^T$.

The value of the acquiring firm before M&A is:

$$V^A(x) = E \int_0^\infty e^{-\rho t} (Q^A X_t) dt = \frac{Q^A x}{\rho - \alpha} \tag{4}$$

The value of the merged firm after M&A is:

$$V^M(x) = E \int_0^\infty e^{-\rho t} (Q^M X_t) dt = \frac{Q^M x}{\rho - \alpha} > \frac{Q^A x}{\rho - \alpha} = V^A(x) \tag{5}$$

The acquiring firm pays the acquired company cash or cash equivalents $P(x)$ to purchase the entire equity of the invested company. The VC owns $\phi$ of the shares of the invested company.

The value of the invested company's shares the VC owns before M&A is:

$$V_{VC}^T(x) = E \int_0^\infty e^{-\rho t} (\phi Q^T X_t) dt = \phi \frac{Q^T x}{\rho - \alpha} \tag{6}$$

It is assumed that the VC use the participating preferred stock to invest, which brings the highest returns to the VC in M&A. The exit earning obtained by the VC by trade sales is $P_{VC}(x)$.

$$P_{VC}(x) = d + \phi [P(x) - d] \tag{7}$$

Where $d$ is fixed income that gets priority in M&A. For using convertible preferred stock or common stock to invest, it is equivalent to the exception where $d=0$ and it is therefore covered.

The VC and the acquiring firm negotiate the merger price to obtain Pareto's effective synergistic value distribution. The merger price is determined by the equilibrium of the Nash bargaining game. We suppose that the negotiation ability of the venture capitalist is $\beta_{VC}$, the negotiation ability of the acquiring firm is $\beta_A = 1 - \beta_{VC}$. We follow Alvarez and Stenbacka (2006), the merger price is the solution to the optimization problem below:

$$\sup_{p^*} \left[ P_{VC}(x) - V_{VC}^T(x) \right]^{\beta_{VC}} \left[ V^M(x) - P(x) - V^A(x) \right]^{\beta_A} \qquad (8)$$

Where $P_{VC}(x) - V_{VC}^T(x)$ is the value increment of the VC through M&A. $V^M(x) - P(x) - V^A(x)$ is the synergy obtained by the acquiring firm after paying the merger price $P(x)$. By solving we find that the Nash bargaining solution is given by

$$P^*(x) = \frac{Q^T x}{\rho - \alpha} - (1 - \beta_{VC})\frac{1-\phi}{\phi} d + \beta_{VC} \left[ \frac{(Q^M - Q^A - Q^T)x}{\rho - \alpha} \right] \qquad (9)$$

The exit earning obtained by the VC is

$$P_{VC}^*(x) = \frac{\phi Q^T x}{\rho - \alpha} + \beta_{VC}(1-\phi)d + \phi\beta_{VC} \left[ \frac{(Q^M - Q^A - Q^T)x}{\rho - \alpha} \right]$$

$$= V_{VC}^T(x) + \beta_{VC}(1-\phi)d + \phi\beta_{VC}\Delta V(x) \qquad (10)$$

The above formula shows that payoffs of the VC in trade sale consists of three parts. The first part $V_{VC}^T(x)$ is the value of the shares held by the VC when the invested company maintains independent operation, and the second part $\beta_{VC}(1-\phi)d$ is the profit from the priority settlement in M&A. The last part $\phi\beta_{VC}\Delta V(x)$ is the synergy benefit the VC sharing.

The return of the VC by trade sale is a stochastic process that is affected by market uncertainty. In order to maximize the return of the VC, it is necessary to choose the optimal exit timing. We set $C$ as the cost of the exit for the VC. We expect to maximize the expected value of exit return for the VC:

$$\sup_{\tau \geq t} E_t \left[ D_n(t,\tau)\left( P_{VC}^*(X_\tau) - C \right) \right] \qquad (11)$$

Where $E_t[\cdot]$ denotes the venture capitalist's conditional expectation at time $t$.

3. Several types of time-inconsistent venture capitalists
3.1 Venture capitalists who only realize critical time point inconsistency

Consider the venture capitalist who only realize critical time point inconsistency while ignoring time flow inconsistency. The assumption about this type of venture capitalist is natural due to the vital importance of VC fund's maturity. Fig.3 shows how this type of venture capitalist think about exit decisions. This optimization problem solving the exit threshold of self 0 is transformed into a two-stage optimization problem that can be solved by backward induction.

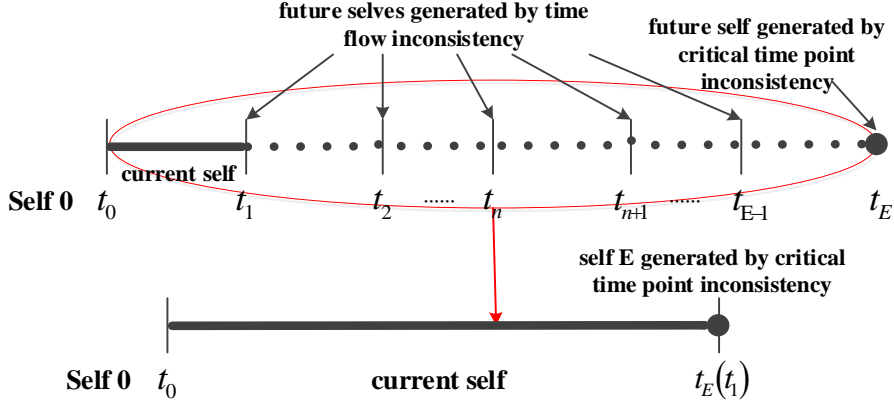

Fig.3 Selves of venture capitalists who only realize critical time point inconsistency

First we analyze the exit timing selection of self $E$. self $E$ faces the case just as the time-consistent venture capitalist. Let $F(x)$ denote self $E$'s exit opportunity value function and $x^*$ be his optimal exit threshold. According to the continuous time Bellman equation $\rho F dt = E(dF)$, $F(x)$ satisfies the differential equation below (see Dixit and Pindyck (1994)):

$$\frac{1}{2}\sigma^2 x^2 F''(x) + \alpha x F'(x) - \rho F(x) = 0 \quad (12)$$

Eq.(12) is solved by using the value-matching condition and smooth-pasting condition.

$$F(x^*) = P_{VC}^*(x^*) - C \quad (13)$$

$$F'(x^*) = \left[P_{VC}^*(x^*) - C\right]' \quad (14)$$

In addition, the implied condition of stochastic process is $F(0) = 0$, and we note that the general solution of $F(x)$ is given by:

$$F(x) = A x^{\beta_1} \quad (15)$$

Where $\beta_1 = \frac{1}{2} - \frac{\alpha}{\sigma^2} + \sqrt{\left(\frac{\alpha}{\sigma^2} - \frac{1}{2}\right)^2 + \frac{2\rho}{\sigma^2}} > 1$.

The exit threshold $x^*$ is given by

$$x^* = \frac{\beta_1(\rho - \alpha)\left[C - \beta_{VC} d(1 - \phi)\right]}{\phi(\beta_1 - 1)\left[Q^T + \beta_{VC}(Q^M - Q^A - Q^T)\right]} \quad (16)$$

To simplify Eq.(16), we assume that $\eta = \dfrac{\phi\left[Q^T + \beta_{VC}(Q^M - Q^A - Q^T)\right]}{\rho - \alpha}$ and $\theta = C - \beta_{VC} d(1 - \phi)$. Thus Eq.(16) can be express to be:

$$x^* = \frac{\beta_1 \theta}{(\beta_1 - 1)\eta} \quad (17)$$

The option value function $F(x)$ before exiting is given by

$$F(x) = \left(\frac{x}{x^*}\right)^{\beta_1} (\eta x^* - \theta) \quad (18)$$

And at the same time we have the exit threshold and the value function of time-consistent venture capitalist. Then we analyze the exit timing selection of self 0. Let $G(x)$ and $x_G$ denote value function and exercise threshold for self 0, respectively. Following Grenadier and Wang (2007) and drawing on the idea of the continuation value function, $G(x)$ solves the differential equation:

$$\frac{1}{2}\sigma^2 x^2 G''(x) + \alpha x G'(x) - \rho G(x) + \lambda_E \left[F^c(x) - G(x)\right] = 0 \quad (19)$$

Where $F^c(x)$ is self 0's continuation value function, upon the arrival of self $E$, which occers at the intensity $\lambda_E$, and $F^c(x) = \delta_p F(x)$. Eq.(19) is solved by using the value-matching and smooth-pasting conditions:

$$G(x_G) = P_{VC}^*(x_G) - C = \eta x_G - \theta \quad (20)$$

$$G'(x_G) = \left[P_{VC}^*(x_G) - C\right]' = \eta \quad (21)$$

We note that the general solution of $G(x)$ is given by:

$$G(x) = A_G x^{\beta_2} + \delta_p F(x) \quad (22)$$

Where $\beta_2 = \frac{1}{2} - \frac{\alpha}{\sigma^2} + \sqrt{\left(\frac{\alpha}{\sigma^2} - \frac{1}{2}\right)^2 + \frac{2(\rho + \lambda_E)}{\sigma^2}} > \beta_1$

The exit exercise threshold and option value function is given in Eq.(23) and Eq.(24):

$$x_G = \frac{1}{\eta(\beta_2 - 1)} \left[\beta_2 \theta + (\beta_2 - \beta_1)\delta_p F(x_G)\right] \quad (23)$$

$$G(x) = \frac{\eta(\beta_1 - 1)}{\beta_2 - \beta_1}(x^* - x_G)\left(\frac{x}{x_G}\right)^{\beta_2} + \delta_p F(x) \quad (24)$$

3.2 Naïve venture capitalists

The naïve venture capitalist, who exactly realizes critical time point inconsistency, mistakenly understands decision criterion of future selves generated by time flow inconsistency. In our model, this type of venture capitalist believes self $n(n = 1 \sim E-1)$ will adopt strategies which are consistent with the current self (self 0). In other words, the discount function of $D_n(t,s)(n = 1 \sim E-1)$ will not update and always equal to $D_0(t,s)$. Therefore, we could view all future selves (not including self $E$) as one self (named self 1 in our model). Fig.4 shows decision criterion of the naïve venture capitalist. It is a three-stage optimalation problem that can be solved by backward induction.

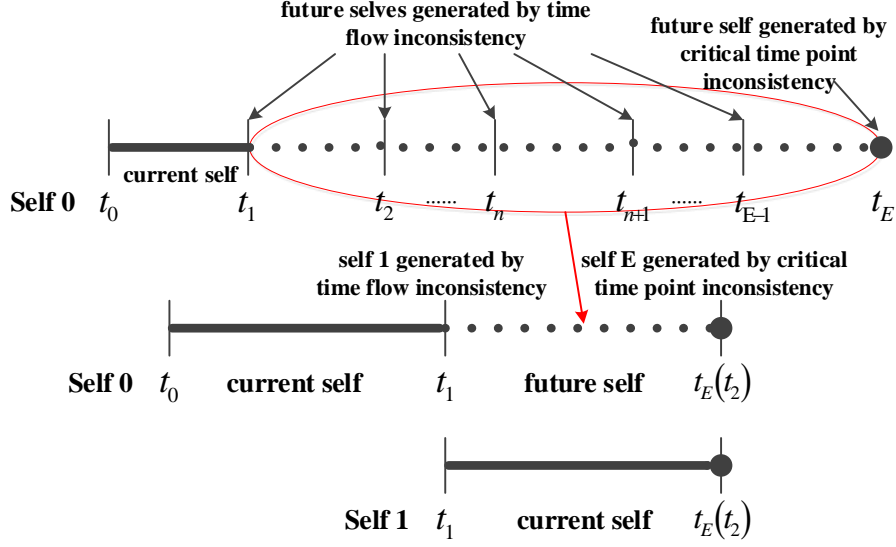

Fig.4 Selves of naïve venture capitalists

First let's consider the optimization problem from self $E$'s viewpoint. As mentioned above, self $E$ faces the case as the time-consistent venture capitalist. Let $N_E(x)$ and $x_{N,E}$ denote self $E$'s value function and exercise threshold, respectively.

$$N_E(x) = F(x) = \left(\frac{x}{x^*}\right)^{\beta_1}(\eta x^* - \theta) \tag{25}$$

$$x_{N,E} = x^* = \frac{\beta_1 \theta}{(\beta_1 - 1)\eta} \tag{26}$$

Then self 1 faces the case that there is only self $E$ in the future. This situation is similar to the type of venture capitalists who only realize critical time point inconsistency. The difference is only the arrival intensity of self $E$. self 1's value function $N_1(x)$ and exercise threshold $x_{N,1}$ is given in Eq.(27) and Eq.(28):

$$N_1(x) = \frac{\eta(\beta_1 - 1)}{\beta_3 - \beta_1}(x^* - x_{N,1})\left(\frac{x}{x_{N,1}}\right)^{\beta_3} + \delta_p F(x) \tag{27}$$

$$x_{N,1} = \frac{1}{\eta(\beta_3 - 1)}\left[\beta_3 \theta + (\beta_3 - \beta_1)\delta_p F(x_{N,1})\right] \tag{28}$$

Where $\beta_3 = \frac{1}{2} - \frac{\alpha}{\sigma^2} + \sqrt{\left(\frac{\alpha}{\sigma^2} - \frac{1}{2}\right)^2 + \frac{2(\rho + \lambda_{pN})}{\sigma^2}}$. Obviously, $\beta_3 > \beta_2 > \beta_1$.

Last self 0 decides on his exercise threshold $x_{N,0}$ in consideration of his future selves' exercise thresholds, as long as his future selves have the opportunity to exercise the exit option. Self 0's continuation value function $N_1^c(x)$ is calculated as follow. If self 1 is alive when his threshold $x_{N,1}$ is reached, then the exit option is exercised, and its payoff to self 0 is $\delta_f \left[ P_{VC}^*(x_{N,1}) - C \right]$. If self 1 dies and self $E$ arrives before $x_{N,1}$ is reached, then self 0's continuation value $N_1^c(x)$ changes into self 1's continuation value $F^c(x)$. Thus $N_1^c(x)$ solves the differential equation:

$$\frac{1}{2}\sigma^2 x^2 N_1^{c\prime\prime}(x) + \alpha x N_1^{c\prime}(x) - \rho N_1^c(x) + \lambda_{pN}\left[F^c(x) - N_1^c(x)\right] = 0 \qquad (29)$$

Where $F^c(x) = \delta_p F(x)$, the value-matching condition is given by:

$$N_1^c(x_{N,1}) = \delta_f \left[P_{VC}^*(x_{N,1}) - C\right] = \delta_f \left(\eta x_{N,1} - \theta\right) \qquad (30)$$

The value-matching condition makes sure the continuity of the continuation value function. We noted that solving $N_1^c(x)$ only requires a boundary condition. To simplify the expression, we assume that $Y = x_{N,1}^{-\beta_3}\left[\delta_f(\eta x_{N,1} - \theta) - \delta_p F(x_{N,1})\right]$. Self 0's continuation value function $N_1^c(x)$ is given by:

$$N_1^c(x) = Y x^{\beta_3} + \delta_p F(x) \qquad (31)$$

Self 0 maximizes his value function $N_0(x)$, by taking his continuation value function $N_1^c(x)$ and choosing his exit threshold $x_{N,0}$. Thus $N_1^c(x)$ solves the differential equation:

$$\frac{1}{2}\sigma^2 x^2 N_0^{\prime\prime}(x) + \alpha x N_0^{\prime}(x) - \rho N_0(x) + \lambda_f \left[N_1^c(x) - N_0(x)\right] = 0 \qquad (32)$$

It is solved by using the value-matching condition and smooth-pasting condition.

$$N_0(x_{N,0}) = P_{VC}^*(x_{N,0}) - C = \eta x_{N,0} - \theta \qquad (33)$$

$$N_0^{\prime}(x_{N,0}) = \left[P_{VC}^*(x_{N,0}) - C\right]^{\prime} = \eta \qquad (34)$$

We guess the general solution of $N_0(x)$ takes the form below and verify the conjecture in Appendix A. Then we discuss the case by case.

$$N_0(x) = \begin{cases} \delta_p F(x) + \varepsilon Y x^{\beta_3} + U_0 x^{\beta_4} & \text{if } \lambda_f \neq \lambda_{pN} \\ \delta_p F(x) + R_1 x^{\beta_4} \log x + R_0 x^{\beta_4} & \text{if } \lambda_f = \lambda_{pN} \end{cases} \qquad (35)$$

Where $\beta_4 = \frac{1}{2} - \frac{\alpha}{\sigma^2} + \sqrt{\left(\frac{\alpha}{\sigma^2} - \frac{1}{2}\right)^2 + \frac{2(\rho + \lambda_f)}{\sigma^2}}$. In general, $\beta_4 \geq \beta_3$.

If $\lambda_f \neq \lambda_{pN}$, we note that $\beta_3 \neq \beta_4$.

Self 0's exercise threshold $x_{N,0}$ is the solution to Eq.(36):

$$x_{N,0} = \frac{\beta_4 \theta}{\eta(\beta_4 - 1)} + \frac{\beta_4 - \beta_1}{\eta(\beta_4 - 1)}\delta_p F(x_{N,0}) + \frac{\beta_4 - \beta_3}{\eta(\beta_4 - 1)}\varepsilon Y x_{N,0}^{\beta_3} \qquad (36)$$

Where $\varepsilon = \frac{\lambda_f}{\lambda_f - \lambda_{pN}}$.

The coefficient before $x^{\beta_4}$ is given by:

$$U_0 = x_{N,0}^{-\beta_4}\left[\eta x_{N,0} - \theta - \delta_p F(x_{N,0}) - \varepsilon Y x_{N,0}^{\beta_3}\right] \tag{37}$$

If $\lambda_f = \lambda_{pN}$, we note that $\beta_3 = \beta_4$ (hereinafter referred to as $\beta_4$).

Self 0's exercise threshold $x_{N,0}$ is the solution to Eq.(38)

$$x_{N,0} = \frac{\beta_4 \theta}{\eta(\beta_4 - 1)} + \frac{\beta_4 - \beta_1}{\eta(\beta_4 - 1)}\delta_p F(x_{N,0}) - \frac{R_1}{\eta(\beta_4 - 1)}x_{N,0}^{\beta_4} \tag{38}$$

Where $R_1 = -\dfrac{\lambda_f Y}{\alpha + \dfrac{1}{2}\sigma^2(2\beta_4 - 1)}$.

The coefficient before $x^{\beta_4}$ is given by:

$$R_0 = x_{N,0}^{-\beta_4}\left[\eta x_{N,0} - \theta - \delta_p F(x_{N,0}) - R_1 x_{N,0}^{\beta_4}\log x_{N,0}\right] \tag{39}$$

3.3 Sophisticated venture capitalists

The sophisticated venture capitalist exactly realize not only critical time point inconsistency, but also time flow inconsistency, as shown in Fig.2. This mean that this type of venture capitalists clearly and correctly understand that all future selves will adopt strategies from their own interests. Therefore, we need to start with self $E$, the last self, and backward derive value function, continuation value function and exit threshold for each self until self 0.

Same as the naïve venture capitalist, we can give self $E$'s value function $S_E(x)$ and exercise threshold $x_{S,E}$, directly.

$$S_E(x) = F(x) = \left(\frac{x}{x^*}\right)^{\beta_1}(\eta x^* - \theta) \tag{40}$$

$$x_{S,E} = x^* = \frac{\beta_1 \theta}{(\beta_1 - 1)\eta} \tag{41}$$

Self $E-1$ faces the case just as self 1 of the naïve venture capitalist. Let $S_{E-1}(x)$ and $x_{S,E-1}$ denote self $E-1$'s value function and exercise threshold, respectively. It is noted that the arrival intensity of self $E$ changed again.

$$S_{E-1}(x) = \frac{\eta(\beta_1 - 1)}{\beta_5 - \beta_1}(x^* - x_{S,E-1})\left(\frac{x}{x_{S,E-1}}\right)^{\beta_5} + \delta_p F(x) \tag{42}$$

$$x_{S,E-1} = \frac{1}{\eta(\beta_5 - 1)}\left[\beta_5 \theta + (\beta_5 - \beta_1)\delta_p F(x_{S,E-1})\right] \tag{43}$$

Where $\beta_5 = \dfrac{1}{2} - \dfrac{\alpha}{\sigma^2} + \sqrt{\left(\dfrac{\alpha}{\sigma^2} - \dfrac{1}{2}\right)^2 + \dfrac{2(\rho + \lambda_{pS})}{\sigma^2}}$. In general, $\beta_4 \geq \beta_5$.

We guess the general solution of $S_{E-2}(x)$ which satisfy the corresponding differential equation takes the form below and solve the exit threshold $x_{S,E-2}$ by the value-matching condition and smooth-pasting condition .

$$S_{E-2}(x) = \begin{cases} \delta_p F(x) + \varphi Z x^{\beta_5} + U_{E-2,0} x^{\beta_4} & \text{if } \lambda_f \neq \lambda_{pS} \\ \delta_p F(x) + R_{E-2,1} x^{\beta_4} \log x + R_{E-2,0} x^{\beta_4} & \text{if } \lambda_f = \lambda_{pS} \end{cases} \quad (44)$$

The proof for the general solution of $S_{E-2}(x)$ is similar to the case of the naïve venture capitalist, so it's not repeated here.

If $\lambda_f \neq \lambda_{pS}$, we note that $\beta_4 \neq \beta_5$.

Self $E-2$'s exercise threshold $x_{S,E-2}$ is the solution to Eq.(45):

$$x_{S,E-2} = \frac{\beta_4 \theta}{\eta(\beta_4 - 1)} + \frac{\beta_4 - \beta_1}{\eta(\beta_4 - 1)} \delta_p F(x_{S,E-2}) + \frac{\beta_4 - \beta_5}{\eta(\beta_4 - 1)} \varphi Z x_{S,E-2}^{\beta_5} \quad (45)$$

Where $\varphi = \dfrac{\lambda_f}{\lambda_f - \lambda_{pS}}$, and $Z = \left[\delta_f(\eta x_{S,E-1} - \theta) - \delta_p F(x_{S,E-1})\right]\left(\dfrac{1}{x_{S,E-1}}\right)^{\beta_5}$.

The coefficient before $x^{\beta_4}$ is given by:

$$U_{E-2,0} = x_{S,E-2}^{-\beta_4}\left[\eta x_{S,E-2} - \theta - \delta_p F(x_{S,E-2}) - \varphi Z x_{S,E-2}^{\beta_5}\right] \quad (46)$$

In summary, for $n \leq E-3$, self $n$'s continuation value function $S_{n+1}^c(x)$ satisfies the differential equation below:

$$\frac{1}{2}\sigma^2 x^2 S_{n+1}^{c\,\prime\prime}(x) + \alpha x S_{n+1}^{c\,\prime}(x) - \rho S_{n+1}^c(x) + \lambda_f\left[S_{n+2}^c(x) - S_{n+1}^c(x)\right] = 0 \quad (47)$$

The value-matching condition is given by

$$S_{n+1}^c(x_{S,n+1}) = \delta_f\left[P_{VC}^*(x_{S,n+1}) - C\right] = \delta_f(\eta x_{S,n+1} - \theta) \quad (48)$$

The solutions for the continuation value functions $S_{n+1}^c(x)$ are presented in Appendix B, and $S_{n+1}^c(x)$ is given by

$$S_{n+1}^c(x) = \delta_p F(x) + \varphi^{E-n-2} Z x^{\beta_5} + \sum_{i=0}^{E-n-3} W_{n+1,i}(\log x)^i x^{\beta_4} \quad (49)$$

Self $n+1$'s value function $S_{n+1}(x)$ satisfies the differential equation

$$\frac{1}{2}\sigma^2 x^2 S_{n+1}^{\prime\prime}(x) + \alpha x S_{n+1}^{\prime}(x) - \rho S_{n+1}(x) + \lambda_f\left[S_{n+2}^c(x) - S_{n+1}(x)\right] = 0 \quad (50)$$

It is solved by using the value-matching and smooth-pasting conditions.

$$S_{n+1}(x_{S,n+1}) = P_{VC}^*(x_{S,n+1}) - C = \eta x_{S,n+1} - \theta \quad (51)$$

$$S_{n+1}^{\prime}(x_{S,n+1}) = \left[P_{VC}^*(x_{S,n+1}) - C\right]' = \eta \quad (52)$$

The solutions for the value functions $S_{n+1}(x)$ are presented in Appendix C, and $S_{n+1}(x)$ is given by

$$S_{n+1}(x) = \delta_p F(x) + \varphi^{E-n-2} Z x^{\beta_5} + \sum_{i=0}^{E-n-3} U_{n+1,i}(\log x)^i x^{\beta_4} \quad (53)$$

The optimal exit threshold for self $n+1$, is the solution to Eq.(54):

$$x_{S,n+1} = \frac{\beta_3 \theta}{\eta(\beta_3-1)} + \frac{\beta_3 - \beta_1}{\eta(\beta_3-1)} \delta_p F(x_{S,n+1}) + \frac{\beta_3 - \beta_4}{\eta(\beta_3-1)} \varphi^{E-n-2} Z x_{S,n+1}^{\beta_5}$$

$$- \frac{k}{\eta(\beta_3-1)} \sum_{k=1}^{E-n-3} U_{n+1,k} \left(\log x_{S,n+1}\right)^{k-1} x_{S,n+1}^{\beta_4} \tag{54}$$

for $n \leq E-3$ and $U_{n+1,k} = W_{n+1,k}$, for $1 \leq k \leq E-n-3$.

If $\lambda_f = \lambda_{pS}$, we note that $\beta_4 = \beta_5$ (hereinafter referred to as $\beta_4$).

Self $E-2$'s exercise threshold $x_{S,E-2}$ is the solution to Eq.(55):

$$x_{S,E-2} = \frac{\beta_4 \theta}{\eta(\beta_4-1)} + \frac{\beta_4 - \beta_1}{\eta(\beta_4-1)} \delta_p F(x_{S,E-2}) - \frac{R_{E-2,1}}{\eta(\beta_4-1)} x_{S,E-2}^{\beta_4} \tag{55}$$

Where $R_{E-2,1} = -\dfrac{\lambda_f Z}{\alpha + \dfrac{1}{2}\sigma^2(2\beta_4-1)}$.

The coefficient before $x^{\beta_4}$ is given by:

$$R_{E-2,0} = x_{S,E-2}^{-\beta_4}\left(\eta x_{S,E-2} - \theta - \delta_p F(x_{S,E-2}) - R_{E-2,1} x_{S,E-2}^{\beta_4} \log x_{S,E-2}\right) \tag{56}$$

In summary, for $n \leq E-3$, self $n$'s continuation value function $S_{n+1}^c(x)$ is given by

$$S_{n+1}^c(x) = \delta_p F(x) + \sum_{i=0}^{E-n-2} V_{n+1,i} (\log x)^i x^{\beta_4} \tag{57}$$

Self $n+1$'s value function $S_{n+1}(x)$ is given by

$$S_{n+1}(x) = \delta_p F(x) + \sum_{i=0}^{E-n-2} R_{n+1,i} (\log x)^i x^{\beta_4} \tag{58}$$

The solutions for the continuation value function $S_{n+1}^c(x)$ and value function $S_{n+1}(x)$ is similar to the case under $\lambda_f \neq \lambda_{pS}$, so it's not repeated here.

The optimal exit threshold for self $n+1$, is the solution to Eq.(59):

$$x_{S,n+1} = \frac{\beta_4 \theta}{\eta(\beta_4-1)} + \frac{\beta_4 - \beta_1}{\eta(\beta_4-1)} \delta_p F(x_{S,n+1}) - \frac{k}{\eta(\beta_4-1)} \sum_{k=1}^{E-n-2} R_{n+1,k} \left(\log x_{S,n+1}\right)^{k-1} x_{S,n+1}^{\beta_4} \tag{59}$$

for $n \leq E-3$ and $R_{n+1,k} = V_{n+1,k}$, for $1 \leq k \leq E-n-2$.

4. Model implications

In this section, we compare the exit threshold of above four types of venture capitalists by examining the properties of the model solutions numerically. The base parameter values are set as $\alpha = 0.02$, $\sigma = 0.2$, $\rho = 0.06$ (following the typical real options models, and making sure that $\rho > \alpha$ for convergence), $Q^M = 1.7$, $Q^A = 1$, $Q^T = 0.5$ (making sure that $Q^M > Q^A + Q^T$ for the positive synergy), $\beta_{VC} = 0.2$, $\phi = 0.4$ (can be anywhere from 0 to 1), and $d = 0.4$, $C = 10$ (can be set freely as long as the model has numerical solutions).

## 4.1 Comparsion of time-consistency and only critical time-inconsistency cases

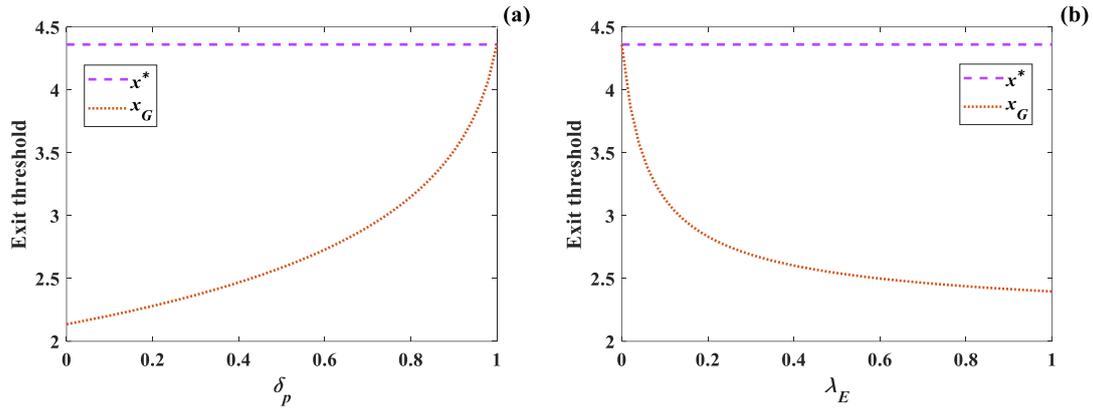

Fig.5 Exit threshold change with respect to reduction factor $\delta_p$ and arrival intensity $\lambda_E$

Fig.5 shows how the exit threshold $x^*$ of time-consistent venture capitalists and $x_G$ of venture capitalists who only realize critical time inconsistency change with parameters $\delta_p$ and $\lambda_E$, respectively. The only critical time-inconsistency case is substantially similar to the model of the typical time-inconsistency papers (such as Grenadier and Wang (2007), Tian (2016), they ignore the diversity of $\delta$ and $\lambda$, and $\delta$ and $\lambda$ both determine the degree of time inconsistency). Therefore as $\delta_p$ increasing form 0 to 1 or $\lambda_E$ decreasing from 1 to 0, the degree of time inconsistency gradually decreases, and the exit threshold $x_G$ gradually approach $x^*$. We separately set $\lambda_E = 1$ in Fig.5 (a) and $\delta_p = 0.3$ in Fig.5 (b).

## 4.2 Comparsion of three types of time-inconsistent venture capitalists

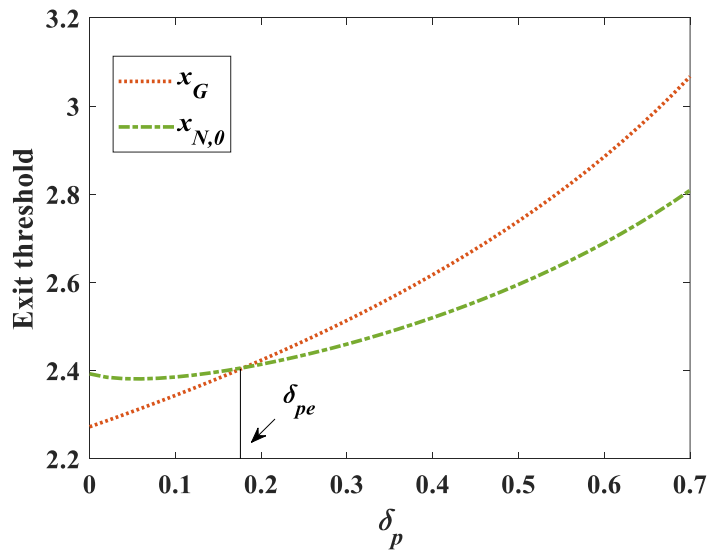

Fig.6 Exit threshold change with reduction factor $\delta_p$ under $\lambda_f = \lambda_{pN} = 1$

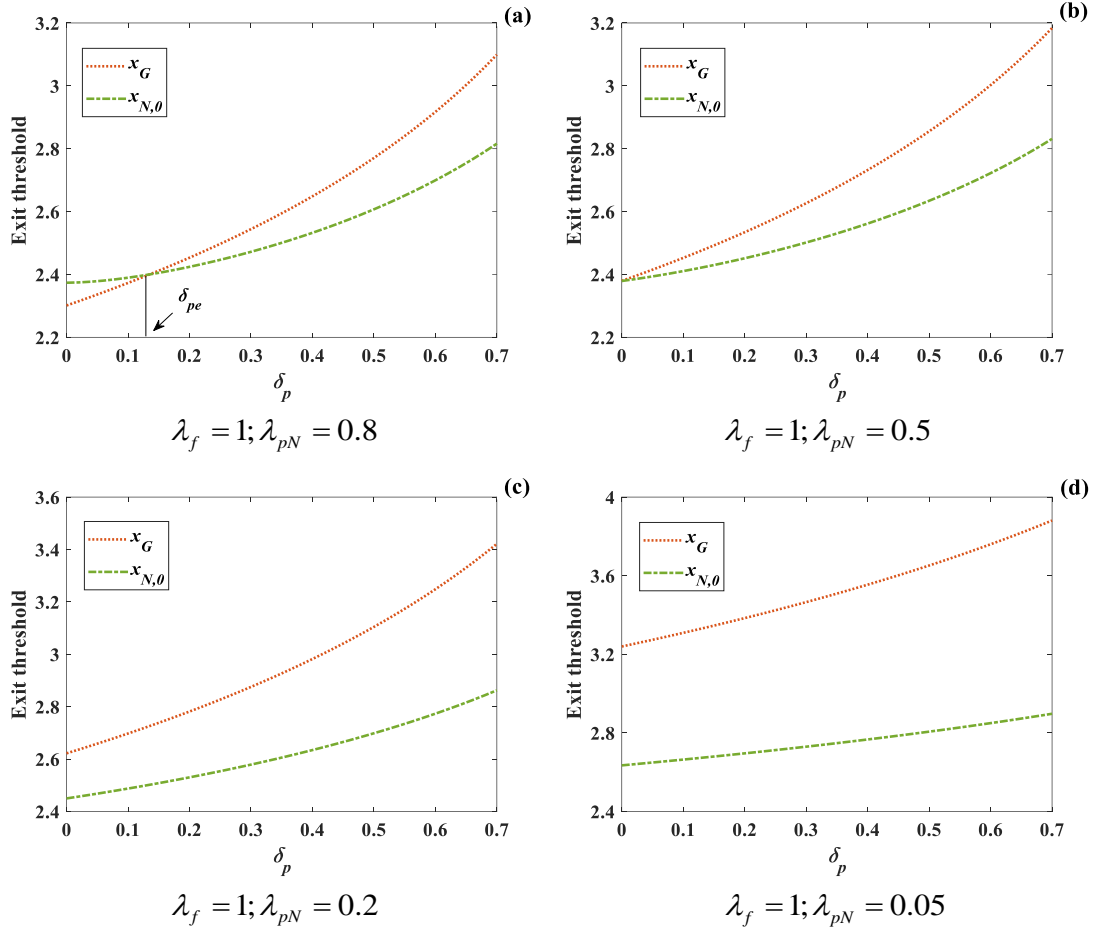

Fig.7 Exit threshold change with reduction factor $\delta_p$ under $\lambda_f \neq \lambda_{pN}$

Fig.6 and Fig.7 demonstate how the exit threshold $x_{N,0}$ of naive venture capitalists and $x_G$ of venture capitalists who only realize critical time inconsistency change with parameter $\delta_p$ under $\lambda_f = \lambda_{pN}$ and $\lambda_f \neq \lambda_{pN}$, respectively. In order to make sure $\delta_f \geq \delta_p$, we set $\delta_f = 0.7$ and $\delta_p$ increase from 0 to 0.7. Considering the randomness of $T_E$, we examine the impact of $\lambda_{pN}$ on the exit threshold of above two kinds of venture capitalists while setting $\lambda_f = 1$ all the time.

Fig.6 shows that $x_G > x_{N,0}$ at $\delta_p > 0.18$, and $x_G < x_{N,0}$ at $\delta_p < 0.18$. Hence we represent the intersection point where $\delta_p = 0.18$ as $\delta_{pe}$. It is noted that when $\delta_p = \delta_f = 0.7$, in other word, we don't distinguish between different $\delta$, our model evolves as the case of Grenadier and Wang (2007). Fig.7 (a), (b), (c), (d) are for the cases where $\lambda_{pN} =$ 0.8, 0.5, 0.2 and 0.05, respectively. Obviously, we find that as $\lambda_{pN}$ decreases, both $x_G$ and $x_{N,0}$ are increasing while $x_G$'s growth rate is greater than $x_{N,0}$'s. When $\lambda_{pN}$ decreases to a certain value, $x_G$ is completely greater than $x_{N,0}$. The intuition is straightforward. As $\lambda_{pN}$ decreases, naïve venture capitalists are more sensitive to the impact of time flow inconsistency, and are more likely to exercise exit options.

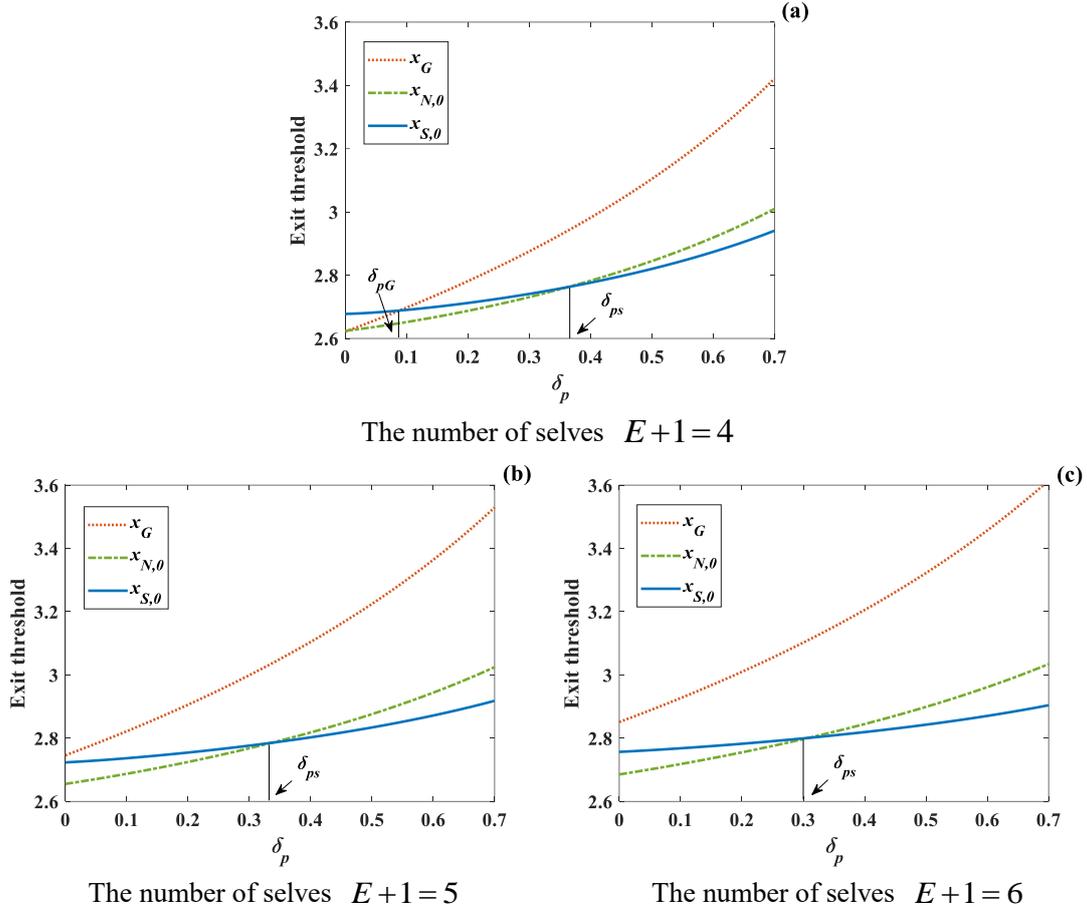

Fig.8 Exit threshold change of three types of venture capitalist with reduction factor $\delta_p$

Fig.8 reveals how exit thresholds $x_G$, $x_{N,0}$ and $x_{S,0}$ change with parameter $\delta_p$ under different numbers of selves which are generated by time flow inconsistency. We set $\lambda_f = \lambda_{pS} = 0.5$ for facilitating the calculation and $\delta_f = 0.7$ as the above case. So the number of selves $E+1$ reflects the randomness of $T_E$. We note that $1/\lambda_{pN} = E/\lambda_f$. Due to the complexity of the calculation, we have given three cases where $E+1$=4, 5 and 6, respectively, which can explain the problem to some extent. It is observed that the growth rate of $x_G$, $x_{N,0}$ and $x_{S,0}$ decreases in turn with the number of selves $E+1$ increasing, as long as the the intersection point $\delta_{pS}$. This result indicates that a gradual increase in the number of selves exerts a multi-level effect on sophisticated venture capitalists resulting in a weakening of future returns, thereby prompting them to speed up the exercise of exit options. When $\delta_p = \delta_f = 0.7$, $x_{N,0} > x_{S,0}$, while our model evolving as the case of Grenadier and Wang (2007).

4.3 The impact of some key factors in VC and M&A on exit threshold of venture capitalists

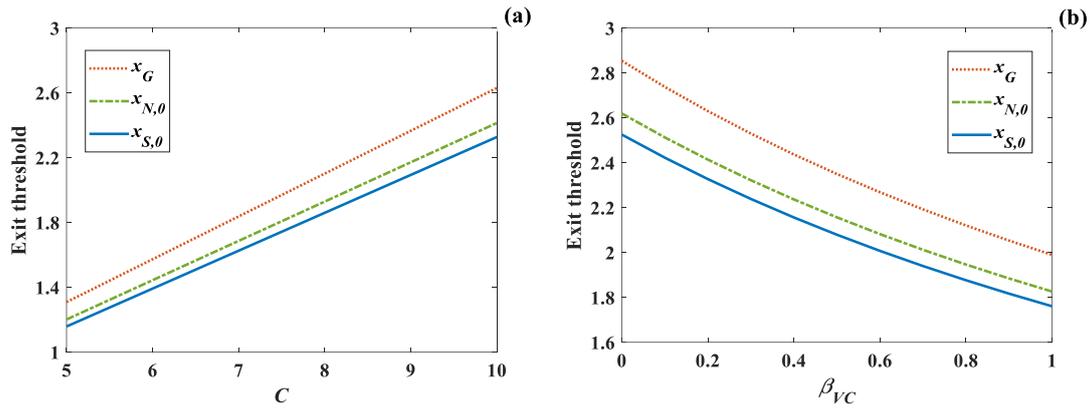

Fig.9 Exit threshold change with respect to exit cost $C$ and negotiation ability $\beta_{VC}$

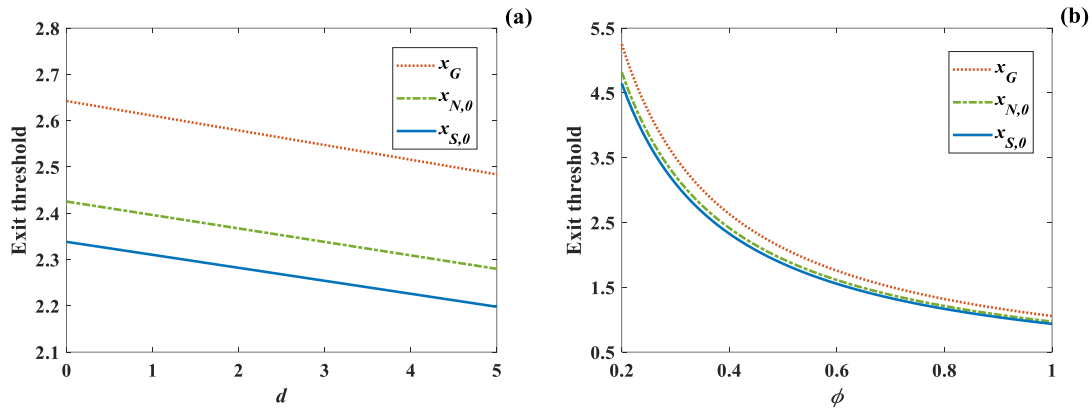

Fig.10 Exit threshold change with respect to fixed income $d$ and equity ratio $\phi$

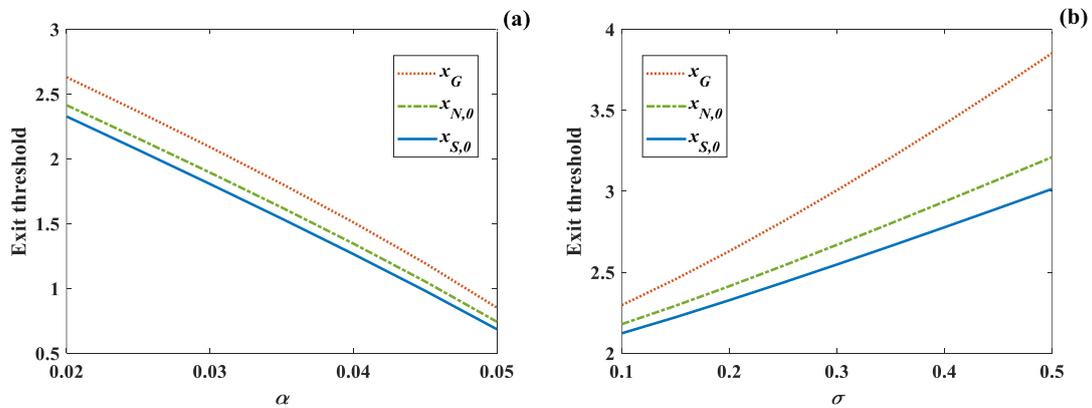

Fig.11 Exit threshold change with respect to expected growth rate $\alpha$ and volatility $\sigma$

In this section, we set $\lambda_f = \lambda_{pS} = 4$ for facilitating the calculation, $E+1=6$, $\delta_p = 0.7$ and $\delta_f = 0.5$ for highlighting the difference between naïve and sophisticated venture capitalists as much as possible. Fig.9 (a) reveals that exit threshold of venture capitalists increases with the increase of exit cost $C$. The increase of exit cost reduces the exit earnings of venture capital, so the optimal choice for venture capitalists is waiting for the invested company to create better performance in the future, which means the higher M&A value. Fig.9 (b) demonstrates that the exit threshold of venture capitalists decreases with the increase of venture capitalists' negotiation ability $\beta_{VC}$. Venture capitalists with strong negotiation ability can share a higher proportion of synergistic gains in trade sales, that means, under the premise of the same performance of the invested company, venture capitalists with strong negotiation ability can obtain the higher exit returns.

Fig.10 (a) shows that the exit threshold of venture capitalists decreases with the increase of fixed income $d$. Through the convertible securities, the VC has the fixed income when it exits. The higher the fixed income, the higher the exit earnings of venture capitalists, so the optimal exit threshold is reduced. Fig.10 (b) indicates that the exit threshold of venture capitalists decreases with the increase of the VC's equity ratio $\phi$. A higher equity ratio allows venture capitalists to obtain higher value compensation in the exit. This means that as the performance of the invested company improves, the opportunity cost of maintaining the company's independent operation instead of exiting is increasing. Therefore, a higher equity ratio encourages venture capitalists to exit as soon as possible.

Fig.11 shows that the exit threshold of venture capitalists decreases with the increase of the invested company's growth rate $\alpha$, while increasing with the increase of the invested company's volatility $\sigma$. The traditional real options literature points out that both growth ($\alpha > 0$) and uncertainty ($\sigma > 0$) can increase the value of waiting, and thereby increase the exit threshold. However, the increase of growth rate $\alpha$ increases the the synergy benefit, and on the other hand decreases the parameter $\beta$. Consequently the exit threshold stems from the interaction of two conflicting factors above, and the frist factor dominates variation trend of exit threshold.

5. Conclusions

5.1 Summary of findings

This paper merges time-inconsistent preferences and the exit decision of venture capital. Hence we get some more realistic and interesting results. In transition economies like China, a large number of venture capital funds have been set up and put into operation. For many of start-up, not prestigious VC funds, obtaining great returns and exiting successfully within the investment period could guarantee the follow-up fund-raising. Consequently, there is a great need to consider time inconsistency when exploring the exit decision of venture capital.

Our paper proposes an optimal exit timing of venture capitalists with time-inconsistent preferences under trade sale. We unify two kinds of time inconsistency in mathematics, respectively from time flow inconsistency mentioned in the previous literature, and critical time point inconsistency caused by the finite lifespan of VC funds. According to venture capitalists' expectations regarding these two kinds of time inconsistency, we distinguish and discuss four types of venture capitalists including (1) time-inconsistent venture capitalists; (2) venture capitalists who only realize critical time point inconsistency; (3) naïve venture capitalists and (4) sophisticated venture capitalists. The results are summarized as follows: (1) all types of time-inconsistent venture capitalists tend to exit earlier than the time-consistent venture capitalist; (2) the longer the expire date are, the more likely time-inconsistent venture capitalists are to delay the exit, but the delay degrees decrease successively (venture capitalist who only realize critical time point inconsistency > naïve venture capitalists > sophisticated venture capitalists). (3) the exit threshold of venture capitalists is positively correlated with exit cost and volatility of the invested company' profit, while it is negatively correlated with negotiation ability of venture capitalists, fixed income in M&A, equity ratio of the VC, and expected growth rate of the invested company' profit.

From a method perspective, our paper inherits the idea of the continuation value function from Grenadier and Wang (2007) to accurately model the exit timing of venture capital. In addition, we extend the time-inconsistency framework to consider the critical time inconsistency whose degree is higher than the degree of time flow inconsistency. The decision-making in this scenario is more realistic for the exit of venture capital. In the end, we notice that there is related literature about time-inconsistent preferences and venture capital contract (see Chapter 17 of Cumming (2012)). This chapter discusses the design of optimal contracts bwtween the entrepreneur and venture capitalist, taking into account self-control problems (for example, procrastination and overconsumption) caused by time inconsistency. And this chapter provides alternative explanations for governance mechanisms of venture capital and some provisions in limited partnership agreements.

5.2 Prospective work

Two possible extensions to this paper are introduced below. Firstly, our paper follows the previous literature and assumes that the discount reduction factor $\delta_f$ is fixed for each selves. Nevertheless, a more realistic assumption would be that the closer to the expire date, the smaller $\delta_f$ is. Secondly, it is assumed that venture capitalists retain control over their capital exit from the invested company. Although in reality venture capital contracts do have provisions to give venture capitalists this right (for example, drag-along rights and tag-along rights, see Bienz and Walz (2010)), entrepreneurs lose the private benefits of control and hence may oppose trade sales. Therefore, the model could be extended to account for the game between entreoreneurs, venture capitalists and acquiring firms.

Appendix A

The proof for the general solution of $N_0(x)$.

The general solution we guess is given by:

$$N_0(x) = \begin{cases} \delta_p F(x) + \varepsilon Y x^{\beta_3} + U_0 x^{\beta_4} & \text{if } \lambda_f \neq \lambda_{pN} \\ \delta_p F(x) + R_1 x^{\beta_4} \log x + R_0 x^{\beta_4} & \text{if } \lambda_f = \lambda_{pN} \end{cases} \quad (60)$$

Take the first and second derivatives of $N_0(x)$ and substitute them into the differential equation:

$$\frac{1}{2}\sigma^2 x^2 N_0''(x) + \alpha x N_0'(x) - \rho N_0(x) + \lambda_f \left[ N_1^c(x) - N_0(x) \right] = 0 \quad (61)$$

If $\lambda_f \neq \lambda_{pN}$, we note that $\beta_3 \neq \beta_4$.

For the $x^{\beta_1}$ term, we have

$$\delta_p F(x) \left[ \frac{1}{2}\sigma^2 \beta_1(\beta_1 - 1) + \alpha \beta_1 - \rho \right] = 0 \quad (62)$$

For the $x^{\beta_3}$ term, we have

$$\frac{1}{2}\sigma^2 \beta_3(\beta_3 - 1) x^{\beta_3}\varepsilon Y + \alpha \beta_3 x^{\beta_3}\varepsilon Y - \rho x^{\beta_3}\varepsilon Y + \lambda_f \left[ x^{\beta_3} Y - x^{\beta_3}\varepsilon Y \right] = 0 \quad (63)$$

Simplify the equation, we obtain

$$\varepsilon = \frac{\lambda_f}{\lambda_f - \lambda_{pN}} \quad (64)$$

For the $x^{\beta_4}$ term, we have

$$U_0 x^{\beta_4} \left[ \frac{1}{2}\sigma^2 \beta_4(\beta_4 - 1) + \alpha \beta_4 - (\rho + \lambda_f) \right] = 0 \quad (65)$$

When $\lambda_f = \lambda_{pN}$, we note that $\beta_3 = \beta_4$.

For the $x^{\beta_1}$ term, we have

$$\delta_P F(x) \left[ \frac{1}{2}\sigma^2 \beta_1(\beta_1 - 1) + \alpha \beta_1 - \rho \right] = 0 \quad (66)$$

For the $x^{\beta_4} \log x$ term, we have

$$R x^{\beta_4} \log x \left[ \frac{1}{2}\sigma^2 \beta_4(\beta_4 - 1) + \alpha \beta_4 - \rho - \lambda_f \right]$$

$$+ x^{\beta_4} \left[ \frac{1}{2}\sigma^2 R(2\beta_4 - 1) + \alpha R + \lambda_f Y \right] = 0 \quad (67)$$

Thus

$$\frac{1}{2}\sigma^2 R(2\beta_4 - 1) + \alpha R + \lambda_f Y = 0 \quad (68)$$

We obtain

$$R = -\frac{\lambda_f Y}{\alpha + \frac{1}{2}\sigma^2(2\beta_4 - 1)} \tag{69}$$

For the $x^{\beta_4}$ term, we have

$$Ax^{\beta_4}\left[\frac{1}{2}\sigma^2\beta_4(\beta_4 - 1) + \alpha\beta_4 - (\rho + \lambda_f)\right] = 0 \tag{70}$$

Appendix B

Solving for the continuation value function $S_{n+1}^c(x)$.

Let $n = N - (j+1)$, for $j = 2, 3, ..., N-1$. Then $S_{n+1}^c(x) = S_{N-j}^c(x)$. We guess the general solution of $S_{N-j}^c(x)$ is given in Eq.(71):

$$S_{N-j}^c(x) = \delta_p\left(\frac{x}{x^*}\right)^{\beta_1}(\eta x^* - \theta) + \varphi^{j-1} Z x^{\beta_5} + \sum_{i=0}^{j-2} W_{N-j,i}(\log x)^i x^{\beta_4} \tag{71}$$

Further, it can be concluded that

$$S_{N-j+1}^c(x) = \delta_p\left(\frac{x}{x^*}\right)^{\beta_1}(\eta x^* - \theta) + \varphi^{j-2} Z x^{\beta_5} + \sum_{i=0}^{j-3} W_{N-(j-1),i}(\log x)^i x^{\beta_4} \tag{72}$$

We substitute the general solution of $S_{N-j}^c(x)$ and $S_{N-j+1}^c(x)$ and the first and second derivatives of $S_{N-j}^c(x)$ into the the differential equation:

$$\frac{1}{2}\sigma^2 x^2 S_{N-j}^{c\,\prime\prime}(x) + \alpha x S_{N-j}^{c\,\prime}(x) - \rho S_{N-j}^c(x) + \lambda_f\left[S_{N-j+1}^c(x) - S_{N-j}^c(x)\right] = 0 \tag{73}$$

Obviously, for the $x^{\beta_1}$ term, Eq.(73) always stand up.

For the $x^{\beta_5}$ term, we have

$$\frac{1}{2}\sigma^2\beta_5(\beta_5 - 1)x^{\beta_5}\varphi^{j-1}Z + \alpha\beta_5 x^{\beta_5}\varphi^{j-1}Z - \rho x^{\beta_5}\varphi^{j-1}Z + \lambda_f\left[x^{\beta_5}\varphi^{j-2}Z - x^{\beta_5}\varphi^{j-1}Z\right] = 0 \tag{74}$$

Simplify the equation, we obtain

$$\varphi = \frac{\lambda_f}{\lambda_f - \lambda_{pS}} \tag{75}$$

For the $x^{\beta_4}$ term, we set the coefficients for each $x^{\beta_4}(\log x)^k$ term to 0 and then we obtain:

$$\frac{\sigma^2}{2}\left[(2\beta_4 - 1)(k+1)W_{N-j,k+1} + (k+2)(k+1)W_{N-j,k+2}\right]$$
$$+ \alpha(k+1)W_{N-j,k+1} + \lambda_f W_{N-j+1,k} = 0, \text{ for } k = 1, 2, ..., j-1 \tag{76}$$

We assume that $\mu = \dfrac{-\beta_4}{\sigma^2\beta_4^2/2 + \rho + \lambda_f}$. The coefficient $W_{N-j,k}$ is given by:

$$W_{N-j,k} = \frac{\lambda_f}{k}\left[\mu\sum_{n=0}^{j-k-2}\left(\frac{\sigma^2\mu}{2}\right)^{n+1} W_{N-j+1,k+n}\prod_{m=0}^{n}(k+m) + \mu W_{N-j+1,k-1}\right] \tag{77}$$

For $k = 1, 2, \ldots, j-2$. $W_{N-j,0}$ is solved by using the value-matching condition.

$$W_{N-j,0} = x_{S,N-j}^{-\beta_4} \left[ \delta_f \left( \eta x_{S,N-j} - \theta \right) - \delta_p \left( \frac{x_{S,N-j}}{x^*} \right)^{\beta_1} \left( \eta x^* - \theta \right) - \varphi^{j-1} Z x_{S,N-j}^{\beta_5} \right]$$

$$- \sum_{i=1}^{j-2} W_{N-j,i} \left( \log x_{S,N-j} \right)^i \tag{78}$$

Appendix C

Solving for the value function $S_{n+1}(x)$

Let $n = N - (j+1)$, for $j = 2, 3, \ldots, N-1$. Then $S_{n+1}(x) = S_{N-j}(x)$. We guess the general solution of $S_{N-j}(x)$ is given in Eq.(79)

$$S_{N-j}(x) = \delta_p \left( \frac{x}{x^*} \right)^{\beta_1} \left( \eta x^* - \theta \right) + \varphi^{j-1} Z x^{\beta_5} + \sum_{i=0}^{j-2} U_{N-j,i} \left( \log x \right)^i x^{\beta_4} \tag{79}$$

We substitute the general solution of $S_{N-j}(x)$ and $S_{N-j+1}(x)$ and the first and second derivatives of $S_{N-j}(x)$ into the the differential equation:

$$\frac{1}{2} \sigma^2 x^2 S_{N-j}''(x) + \alpha x S_{N-j}'(x) - \rho S_{N-j}(x) + \lambda_f \left[ S_{N-j+1}^c(x) - S_{N-j}(x) \right] = 0 \tag{80}$$

Obviously, for the $x^{\beta_1}$, $x^{\beta_3}$ term, we reach the same conclusion as Appendix B.

For the $x^{\beta_4}$ term, we set the coefficients for each $x^{\beta_4} \left( \log x \right)^k$ term to 0 and then we obtain:

$$\frac{\sigma^2}{2} \left[ (2\beta_4 - 1)(k+1) U_{N-j,k+1} + (k+2)(k+1) U_{N-j,k+2} \right]$$
$$+ \alpha (k+1) U_{N-j,k+1} + \lambda_f W_{N-j+1,k} = 0, \text{ for } k = 0, 1, \ldots, j-2 \tag{81}$$

The coefficient $U_{N-j,k}$ is given by:

$$U_{N-j,k} = W_{N-j,k} \tag{82}$$

For $k = 1, 2, \ldots, j-1$. $U_{N-j,0}$ is solved by using the value-matching condition.

$$U_{N-j,0} = x_{S,N-j}^{-\beta_4} \left[ \eta x_{S,N-j} - \theta - \delta_p \left( \frac{x_{S,N-j}}{x^*} \right)^{\beta_1} \left( \eta x^* - \theta \right) - \varphi^{j-1} Z x_{S,N-j}^{\beta_5} \right]$$

$$- \sum_{i=1}^{j-2} U_{N-j,i} \left( \log x_{S,N-j} \right)^i \tag{83}$$